\documentclass[10pt]{wlscirep}
\usepackage[utf8]{inputenc}
\usepackage[T1]{fontenc}

\usepackage{graphicx}
\usepackage{indentfirst}
\usepackage{latexsym}
\usepackage{multirow}
\usepackage{epsfig}
\usepackage{multirow}
\usepackage{hyperref}
\usepackage{color}
\usepackage{amssymb}
\usepackage{amsfonts}
\usepackage{amsmath}
\usepackage{bm}
\usepackage{mathrsfs}

\def\l{\lambda}

\newcommand{\mb}[1]{\mbox{\boldmath $#1$}}
\newcommand{\be}{\begin{equation}}
\newcommand{\ee}{\end{equation}}
\newcommand{\beq}{\begin{equation*}}
\newcommand{\eeq}{\end{equation*}}

\title{On ab initio-based, free and closed-form expressions for gravitational waves}

\author[1]{Manuel Tiglio}
\author[1]{Aarón Villanueva}
\affil[1]{Universidad Nacional de C\'ordoba, Facultad de Matem\'atica, Astronom\'ia, F\'isica y Computaci\'on, C\'ordoba, 5000, Argentina}

\affil[*]{mtiglio@unc.edu.ar}

\affil[+]{The authors contributed equally to this work}


\begin{abstract}
We introduce a new approach for finding {\em high accuracy, free and closed-form expressions} for the gravitational waves emitted by binary 
black hole collisions from {\em ab initio models}. 
More precisely, our expressions are built from numerical surrogate models based on supercomputer simulations of the Einstein equations, which have been shown to be essentially indistinguishable from each other. Distinct aspects of  our approach are that: i) representations of the gravitational waves can be {\em explicitly} written in a few lines, ii) these representations are free-form yet still fast to search for and validate and iii) there are no underlying physical approximations in the underlying model. 
The key strategy is combining techniques from Artificial Intelligence and Reduced Order Modeling for parameterized systems. 
Namely, symbolic regression through genetic programming combined with sparse representations in parameter space and the time domain using Reduced Basis and the Empirical Interpolation Method enabling fast free-form symbolic searches and large-scale a posteriori validations. As a proof of concept we present our results for the collision of two black holes, initially without spin, and with an initial separation corresponding to $25-31$ gravitational wave cycles before merger. The minimum overlap, compared to ground truth solutions, is $99\%$. That is, $1\%$ difference between our closed-form expressions and supercomputer simulations; this is considered for gravitational (GW) science more than the minimum required due to experimental numerical errors which otherwise dominate. This paper aims to contribute to the field of GWs in particular and Artificial Intelligence in general. 
\end{abstract}
\begin{document}

\flushbottom
\maketitle
\thispagestyle{empty}

\section*{Introduction}

The surge of direct detections of the gravitational waves (GWs) emitted by the collision of two black holes, neutron stars, and mixed pairs through laser interferometer laboratories \cite{LIGO-detection-papers} stresses even more the need for fast evaluation of the GWs emitted by these processes, as predicted by Einstein's theory of gravity (or alternative ones \cite{Cornish:2011ys}).  High accuracy numerical simulations of the Einstein equations are the gold standard for these predictions. The problem is that they are computationally very expensive. As an example, if one considers the case of two black holes, initially far away and in quasi-circular orbit, the usual rationale is that due to the no-hair theorem each black hole can be uniquely described by its mass and spin. Which results in eight degrees of freedom. The total time of computation (wall time) elapsed for each of these simulations depends, among other factors, on the initial separation of the two black holes, resolution, etc. For the sake of definiteness, let's say that each simulation takes $10,000$ hours of wall time, which is a lower bound for cases of interest. Assuming that one samples each parameter dimension with, say, $100$ points, this gives $10^{16}$ years of computing time, which is orders of magnitude larger than the age of the universe. Even using the top 10 supercomputers \cite{top500} this time would be reduced to $\sim 10^8$ years. Worse, for Bayesian parameter estimation, a catalog/bank of templates cannot be constructed a priori since each new waveform needs to be computed on demand without a priori knowledge of which ones those will be \cite{PhysRevD.94.044031}. All these are consequences of the curse of dimensionality combined with an already computationally very expensive High Performance Computing problem (numerically solving the Einstein field equations) even for any fixed parameter tuple. 

This bottleneck cannot be overcome through software optimization or specialized hardware (such as Graphics Processing Units, for example), which gave rise to the introduction of Phenomenological \cite{Ajith:2007qp} and Effective One Body (EOB) \cite{Buonanno:1998gg} models for the prediction of GWs. These do not correspond to solutions of the Einstein field equations but, instead, physically inspired fits and/or ``stiches'' of approximate models for binary systems. We will not review these approaches here -- except some remarks in the final section -- since our take is to represent the emitted GWs using Einstein's theory of gravity without any physical approximation.

The GWs emmitted by binary systems can be seen as a parameterized problem, the parameters being for example the masses and spins of the binary components, initial separation, equations of state in the case of neutron stars, and so on. For each value of the parameter tuple, the GW consists of two degrees of polarization which can be encapsulated in a single complex-valued time (or, equivalently, frequency) series. 

Over the last decade surrogate models for GWs based on modern frameworks such as Reduced Basis (RB) and the Empirical Interpolation Method (EIM) \cite{Rifat:2019ltp, Varma:2018mmi, Blackman:2015pia, Caudill2012, PhysRevD.95.104023, PhysRevResearch.1.033015, Blackman:2015pia, Field:2013cfa, Field:2012if, Field:2011mf, Canizares:2014fya} have been developed. For a recent and detailed review and any doubts in what follows in this article regarding reduced order and surrogate modeling see~Ref.\cite{tiglio2021reduced}. The RB framework provides a sparse representation in the parameter domain while the EIM does so in the dual one, here being time. Both  make use of the smooth dependence on parameter variation to achieve fast -- usually exponential in the case of GWs -- convergence of the representation with respect to the number of basis elements (which by construction equals the number of EIM time nodes). The resulting surrogate models predict GWs which are essentially indistinguishable from numerical relativity (NR) simulations, but with a speedup of evaluation of around eight orders of magnitude \cite{Blackman:2015pia}, being evaluated in less than a second on a standard laptop instead of using supercomputers as in NR.  A relatively small number of NR simulations are still needed in the {\it offline} (training) stage, though, but with a fast, highly accurate, surrogate, predictive model to evaluate for any parameter and time in the intervals considered in the {\it online} stage. This is referred to as an offline-online decomposition strategy.
 
In this article we build upon these efforts for what we consider a natural next step: a methodology for finding high accuracy free and closed-form representations of GWs, as opposed to {\em numerical} surrogates. As a proof of concept we present our results for the collision of two black holes, initially without spin, and with an initial separation corresponding to about $25-31$ GW cycles before merger. The minimum overlap obtained, compared to ground truth solutions, is $99\%$. That is, $1\%$ difference between our closed-form expressions and supercomputer simulations.
 
 \section*{Method}

\subsection*{Reduced Order and Surrogate Models}
In general, surrogate models for GWs have followed reduced order modeling (ROM) for parameterized systems and/or ``standard'' machine learning regression techniques \cite{10.1088/1361-6382/ab693b}. Here we focus on the former, we will not delve into reviewing them but instead, as mentioned, refer to \cite{tiglio2021reduced}. In this work we focus on RB \cite{Jan-RB, quarteroni2015reduced} and the EIM \cite{Maday_2009, Barrault2004667,sorensen}. Briefly, RB collocates parameter points according to their relevance, which are used to build a hierarchical, nearly optimal basis in a rigorous mathematical sense with respect to the Kolmogorov n-width \cite{Pinkus, Binev10convergencerates}. The framework of RB takes advantage of any regularity with respect to parameter variation to achieve fast convergence in the accuracy of the representation with the number of basis elements; it is usually referred to as an {\it application-based spectral expansion}. In fact, in the case of gravitational waves it can be easily argued that the parameter dependence is smooth ($C^{\infty}$) and RB has been shown to achieve asymptotic exponential convergence \cite{Field:2011mf, Caudill2012, Field:2012if, Field:2013cfa, Blackman:2014maa, Rifat:2019ltp, Varma:2018mmi} in practice, as expected with any spectral-type method. Similarly, the EIM achieves a subsampling in the space dual to that one of parameters (time in the case here considered) which is also nearly optimal. For a detailed discussion on the optimality of the EIM see~\cite{Villanueva:2020ixh} and references therein. On top of that, high accuracy predictive models (prediction as opposed to projection) can be built once one has a reduced basis and an empirical interpolant \cite{Field:2013cfa}. These predictive models are essentially indistinguishable from numerical relativity supercomputer simulations of the Einstein equations but can be evaluated in less than a second on a laptop. The availability of high accuracy, fast to evaluate, numerical predictive models and a sparse subsampling in time are key components upon we build on in the approach presented in this article for finding and validating ab initio symbolic expressions for GWs.

\subsection*{Symbolic Regression using Genetic Programming}
Symbolic regression (SR) is the general procedure of finding closed-form expressions representing data. Unlike more conventional regression approaches, when SR is free-form it means precisely that: no specific form is specified. This eliminates any possible bias or human knowledge gap when postulating specific forms to fit for; it also contemplates completely data-driven cases, in which there is no underlying fundamental model or if there is one it is (still) unknown~\cite{Schmidt03042009}. 

Genetic programming (GP) is, in brief, an area of  Artificial Intelligence (AI) whose goal is the evolution of programs or tasks through computer means. The techniques of GP emulate those of Nature; that is, algorithms are modeled following the process of natural evolution. A thorough book on  GP is \cite{Koza:1992:GPP:138936}, and a shorter {\it field guide} \cite{AFieldGuide}. 

Unlike perhaps more conventional Machine Learning (ML) deterministic regression approaches, GP-SR uses genetic algorithms to find free, closed-form expressions, either algebraic or differential. GP-SR can be described through the following general tree-structured algorithm tracing genetic programming principles:\\
\rule{8cm}{0.3mm}\\
\noindent1. Create stochastically an initial population of programs (e.g. mathematical expressions and operations);\\
2. {\bf Repeat}\\
3. Execute each program and compute their quality or {\it fitness};\\
4. Select one or two programs from the population with a probability based on their fitness to participate in genetic operations;\\
5. Create new programs through the application of genetic operations (e.g. mutation or crossover);\\
6. {\bf Until} an acceptable solution is found or some other stopping condition is met;\\
7. {\bf Return} the best-so-far individual/s.\\
\rule{8cm}{0.3mm}\\

GP-SR algorithms do not find a unique representation of data but a number of them with different levels of complexity (roughly speaking and for simplicity, cost of evaluation) and fitness with respect to training and validation sets. So, depending on the criteria used to find expressions via SR, the final symbolic forms can be shorter or larger with variable accuracy. In this work we prioritize accuracy. We used Eureqa \cite{eureqa} for GP-SR, although there are open source alternatives such as {\it gplearn}~\cite{gplearn} and {\it Glyph}~\cite{glyph}. As a fitness criteria we used the {\it Hybrid Correlation} and the {\it $R^2$-Goodness-of-fit} metric errors to fit symbolic models for amplitude and phase, respectively. Our contribution is not actually on the area of GP-SR but on how to address its high computational cost by means of modern ROM and show that it actually works through the case study of the gravitational waves emitted by the collision of two black holes (a highly non-trivial system to model). For the details on the GP side of Eureqa see the original works of Lipson, Schmidt and Bongard~\cite{lipson2005,Schmidt03042009,Bongard9943,Schmidt2006,lipson2008,Schmidt2010,Schmidt2011}.
~\\
\noindent{\bf An example: a robot discovering Newton's second law}.
As an example of the power of GP-SR we present results for the following system: the simple pendulum. In polar coordinates, Newton's second law is

\be\label{eq:pend}
\ddot\theta=-\lambda\sin(\theta)\,,
\ee
where $\lambda=g/l$, $g$ is the gravitational acceleration and $l$ the length of the pendulum. The variable $\theta$ represents the angle with respect to the point of stable equilibrium (the pendulum at rest). We set $\lambda=0.5 s^{-2}$ and performed GP-SR on a dataset with initial conditions $\theta(t=0) = \pi/2$ and $\dot\theta(t=0) = 0$. The time interval for the training set was set to be $[0,22]s$ with grid size $\Delta t = 0.1s$, which roughly corresponds to 2 cycles of the pendulum. That is, a single stream of data was used for training. We solved the above ordinary differential equation (ODE) with the {\tt integrate.odeint} solver from the Scipy Python library~\cite{scipy}, and used the resulting data (with intrinsic noise, due to the numerical errors of the ODE solver) to find symbolic expressions for the underlying differential equation, searching for expressions of the form 
$$
\ddot\theta=f(\dot\theta,\theta,t)\,.
$$
The representation with highest fitness -- found in the order of seconds -- was {\em exactly, not a numerical approximation of}, Newton's second law for this system, Eq. (\ref{eq:pend}). Furthermore, for initial conditions close to the stable equilibrium state, one of the symbolic expressions found was {\em exactly} the harmonic oscillator equation.

Although the following conclusion could be somewhat debatable, the point is that a robot could find Newton's fundamental second law in seconds. One could argue that this was for a particular system but, though not presented in these terms, this is the process of scientific induction. In data science (DS), ML or AI, this process is called {\it validation}, whereas in physics it is called {\it verification} (as in verifying Newton's or Einstein's theory of gravity).  In fact, with more computational power, the authors of Eureqa remarkably rediscovered Newton's second law for the double pendulum \cite{Schmidt03042009}, which is known in physics as a classical example of a chaotic system. 

\subsection*{Complexity: Searches and Validations}

Genetic programming symbolic regression algorithms are known for their scalability issues with the amount and dimensionality of data \cite{ONeill2010, 6557774}. Although it is not unusual for deterministic ML algorithms to suffer from the curse of dimensionality, scaling in GP-SR is nevertheless an issue. For our domain application, as described later on in this article, it resulted in no signs of convergence whatsoever after $\mathcal{O}(10^3-10^4)$ core hours (total number of hours with all cores running in parallel). The problem was not so much the number of hours but the lack of any progress in the fitness. 

Our approach to overcome this problem is intuitive and conceptually simple: it uses a set of sparse data for fast training and later high-accuracy surrogate models for large scale validations. Here we use the RB and EIM frameworks combined with the surrogate approach developed in Ref.~\cite{Field:2013cfa}.  If, as in our case, the surrogate models are essentially indistinguishable from supercomputer simulations of the Einstein equations, they can be considered as ground truth solutions, with the advantage of very fast evaluations. The steps of validation for building surrogate models based on RB and the EIM are described in \cite{Field:2011mf,Field:2013cfa,Caudill2012,Blackman:2015pia}. So here we focus on the ones related to GP-SR. In this processes we used a fraction of our catalog for training and another one for validation so as to avoid overfitting; typically we used $50 \% $ for each (training and validation). We then compared the symbolic time series with the ground truth solutions using dense sampling in parameter and time. This validation instance was achieved by computing the overlap integral $S(h^{1},h^{2})(q)$ between surrogate and symbolic normalized waveforms in the time domain, defined as
\be
S(h^{1}, h^{2}):= \operatorname{Re}\langle h^{1}|h^{2} \rangle=1-\frac{1}{2}||h^{1}-h^{2}||^2\,, \label{eq:overlap}
\ee
where
$$
\langle h^{1}|h^{2} \rangle:=\int_{t_{\tt min}}^{t_{\tt max}}dt\,\bar{h}^{1}(t)\,h^{2}(t)\,,
$$
and $\bar{h}^{1}$ stands for the complex conjugate of $h^1$. The overlap $S$ gives a measure of the match between two waveforms and is commonly used in GW science. For training issues we have normalized time by a factor of $1,000$; this has to be taken account for.

\section*{Results}

\subsection*{Gravitational Waves Setup}

As a proof of concept we tackled the problem of two black holes initially in quasi-circular orbit and without spin, for about $25-31$ GW cycles before merger. More precisely, for the time interval $t\in[-2,750: 100]M$, where $M$ is the total mass of the system and the waveforms are aligned so that $t=0$ corresponds to the peak of their amplitudes, which is around the time the two black holes merge. Due to the scale invariance of General Relativity, the only free parameter then is the mass ratio 
\be
q:=m_1/m_2 \, , \label{eq:mass-ratio}
\ee
here chosen to be in the range $q\in[1,10]$, with $m_i$ the mass of each black hole. Furthermore,  for definiteness we restrict our discussion to the dominant multipolar mode, the $\ell =m = 2$ one. 

A surrogate for this problem was constructed in Ref.~\cite{Blackman:2015pia}, and is publicly available as part of the {\it GWSurrogate} Python package \cite{GWSurrogate}.  This surrogate model consists of only $22$ basis elements and, by construction, only $22$ EIM time nodes. These are the only pieces of information needed to predict with high accuracy {\em any} waveform in the parameter and time domains considered. The surrogate model can be considered -- and we do -- as ground truth solutions of the EE, since it was shown in \cite{Blackman:2015pia} that it is essentially indistinguishable from NR simulations up to the numerical errors of the latter, performed by SpEC \cite{spec}, the most accurate code  in the  NR community to date for modeling GWs from sources without shocks (such as binary black holes). 

The two polarizations of GWs can be encoded in a single complex parameterized function, 
$$
h(t,\mb\l):=h_+(t,\mb\l)+i\,h_{\times}(t,\mb\l) \, , 
$$
where $\mb\l$ represents a tuple in parameter dimension; here, it corresponds to the mass ratio $q$ defined in Eq.~(\ref{eq:mass-ratio}).  The waveforms for the collision of two black holes in initial quasi-circular orbit have an apparent complexity, but they are simply oscillatory functions with an increasing amplitude until the time of coalescence, followed by a damped exponentially decaying profile, the quasinormal modes of the final black hole. It is therefore convenient to consider the amplitude $A(t,\mb\l)$ and phase $\phi(t,\mb\l)$ separately, 
\be
h(t,\mb\l) := A (t,\mb\l) \times e^{i \phi (t,\mb\l)} \, , \label{complex}
\ee
find closed-form expressions for them, later reconstruct the symbolic waves and compare them with their ground truth counterparts for a large number of validation cases. 

For both amplitude and phase our symbolic expressions have an $R^2$-goodness-of-fit of at least $R^2 \sim0.999$ with respect to the validation members of the catalog used in the symbolic regression searches. We discuss more thorough and large-scale validation results below. 

For both phase and amplitude our dictionary is composed of the following functions and operations:
\begin{eqnarray*}
&& \tt{Constant}, \tt{Addition}, \tt{Substraction}, \tt{Multiplication}, \tt{Division}, \tt{Sine}, \tt{Cosine}, \tt{Tangent}, \tt{Exponential}, \\
&& \tt{Natural\,Logarithmic}, \tt{Power}, \tt{Square\,Root}, \tt{Gaussian\,Function}, \tt{Hyperbolic\,Tangent}, \tt{Hyperbolic\,Sine}, \\
&& \tt{Hyperbolic\,Cosine}, \tt{Arcsine}, \tt{Arccosine}, \tt{Arctangent}, \tt{Two-Argument\, Arctangent}, \\
&& \tt{Inverse\,Hyperbolic\,Sine},  \tt{Inverse\,Hyperbolic\,Cosine}, \tt{Inverse\,Hyperbolic\, Tangent}.
\end{eqnarray*}


\subsection*{Amplitude}
\begin{figure}[!h]
\begin{center}
\includegraphics[width=0.6\linewidth]{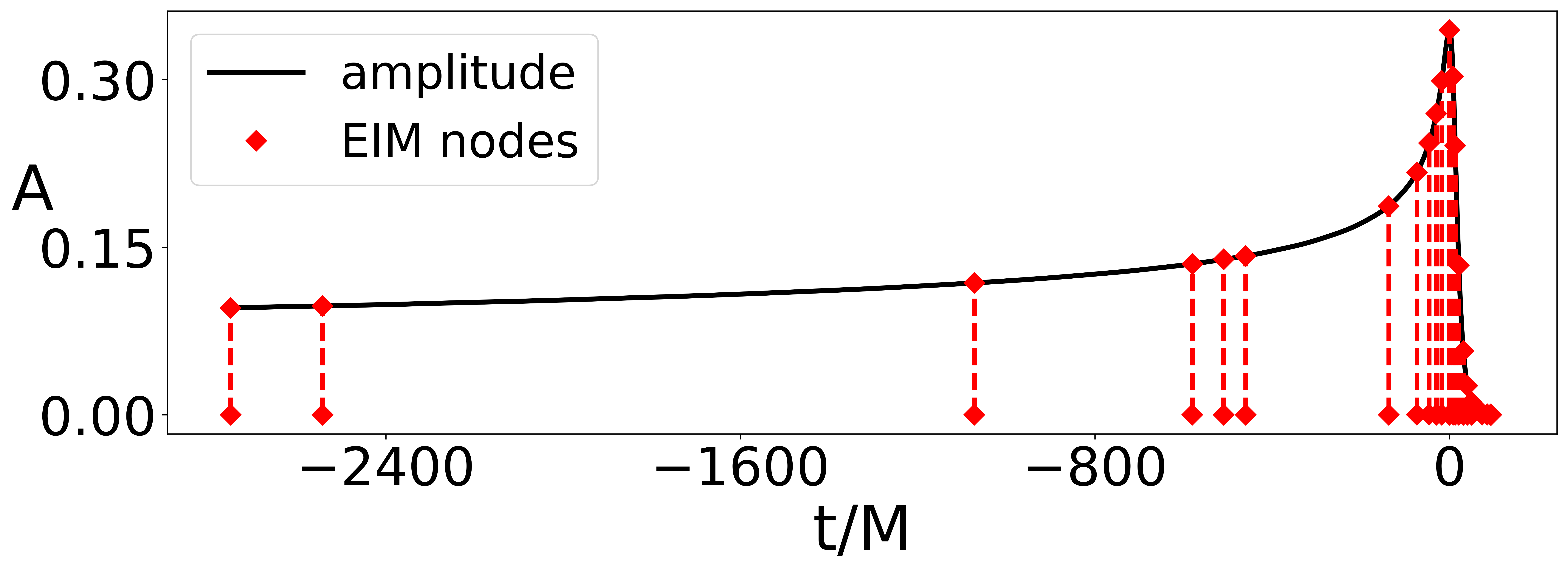}
\caption{Amplitude for the surrogate waveform $q=2$; the red diamonds denote the EIM time nodes, which are by construction the same for all $q\in [1,10]$. Their adaptive nature, leading to a resolution improvement around the peak of the amplitude, can be noticed.}
\label{fig:EIM}
\end{center}
\end{figure}

In our experience, naively sampling both in parameter (mass ratio $q$) and physical dimension (time) resulted in days or weeks of no convergence while searching for symbolic expressions for the amplitude of the GWs. The reason for this is the need to resolve with high accuracy the region around the peak of the amplitude (see Fig.~\ref{fig:EIM}), for which we tried using a dense sampling in time, which led to substantial computation with no signs of convergence. 

One could attempt to manually collocate time nodes where needed. This approach is not only tedious but also not guaranteed to work. Instead, here we resorted to subsampling in time using only the $22$ EIM nodes, shown in Fig.\ref{fig:EIM}, and $90$ equally spaced values in the mass ratio. The rationale for this approach is that the EIM time nodes are the only relevant ones for recovering the whole time series and thus the only representative ones; this intuition proved to be correct, as we discuss below. 
Using {\em only} the EIM nodes in a few minutes we were able to find the following closed-form expression for all $q \in [1,10]$ (we discuss validation using a dense set of time nodes below):  
\be\label{eq:ampq1}
\begin{split}
A(t,q)=	&\{a_1\,\text{gauss}(\text{atan2}(t, a_2 - t))\}/ \{a_3 + q - t - a_4\,\text{gauss}(\text{atan2}(a_5, q) - a_6\,t) \text{gauss}(\text{atan2}(t, a_7 - t -a_8\,t\,q))\} -a_9\,,
\end{split}
\ee
where $\text{gauss}(x):= e^{-x^2}$, $\text{atan2}(x,y)$ is the arctangent of two parameters and
\begin{equation*}
\begin{split}
&a_1=1.37502533181183\,,\,a_2=0.0409895367586908\,,a_3=3.40043449934568\,,\,a_4=1.86434379599601\,,\\
&a_5=1.1446516014466\,,\,a_6=1.49686180948812\,,a_7=0.0250835926883564\,,\,a_8= 0.108134472792241\,,\\
&a_9=0.00178301085458751\,.
\end{split}
\end{equation*}

\begin{figure}[!h]
\begin{center}
\includegraphics[width=0.8\linewidth]{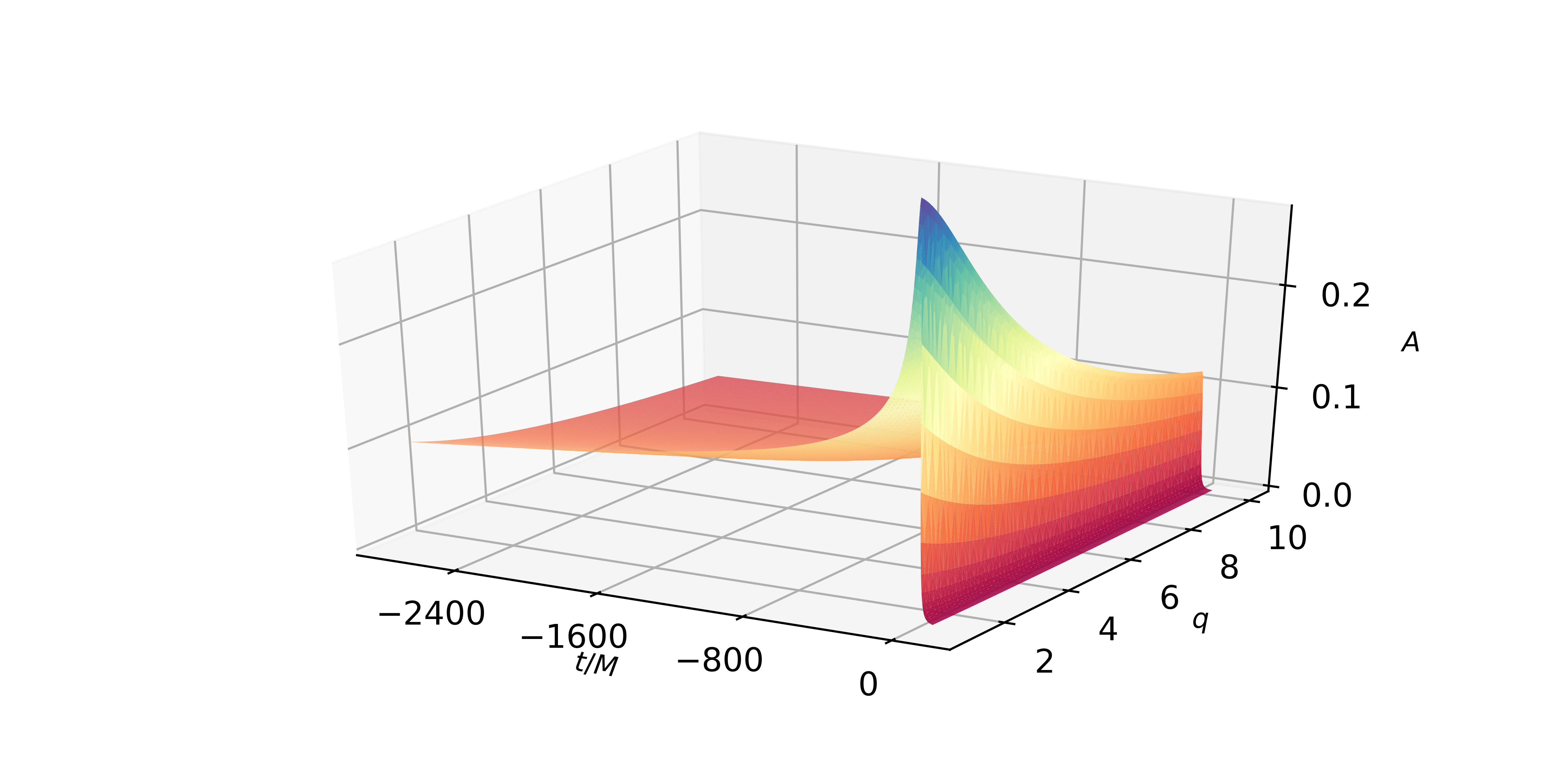}
\caption{Symbolic amplitude~(\ref{eq:ampq1}) as a function of the mass ratio $q=m_1/m_2$ and time. }
\label{fig:amp}
\end{center}
\end{figure}

Figure \ref{fig:amp} shows the symbolic amplitude~(\ref{eq:ampq1}) as a function of the mass ratio $q$ and time. The ground truth values are not shown because they are indistinguishable by eye. 

\begin{figure}[!h]
\begin{center}
\includegraphics[width=0.6\linewidth]{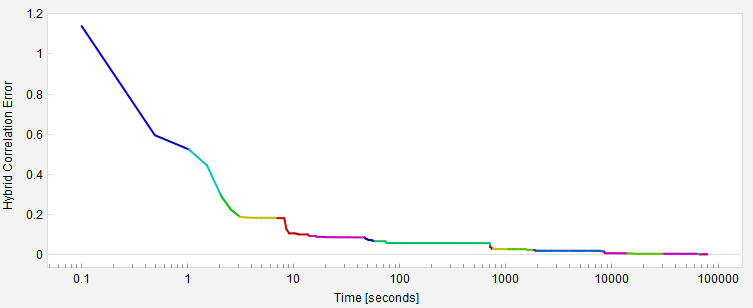}
\caption{Error curve in time for symbolic regression corresponding to the amplitude. Note the error in a few seconds reaches a plateau. We stopped our search when the symbolic waveforms reached a $99\%$ overlap with respect to the surrogate model.}
\label{fig:error_amp}
\end{center}
\end{figure}

As a mode of illustration, we show in Figure~\ref{fig:error_amp} the error curve as a function of time for the amplitude. Although the complete range of convergence included several hours, the plateau of the curve is achieved in the order of minutes, dedicating most of the time to fine tuning of parameters for improving the accuracy of the model. There are several  reasons for the formation of this plateau, for example the finiteness of the dictionary, which serves as a constraint for the search space of functions; and the penalization for large formulas (high complexity) in the GP algorithm, which prevents from finding extremely high complex functions with little or no gain in accuracy. One could say that an important reason for the plateau is the stagnation in local optima, but actually the algorithm softens this problem by implementing a protocol that allows to promote population diversity in the evolutionary search without impacting the fitness performance. For details, see~\cite{Schmidt2006,lipson2008}. 

\subsection*{Phase}
\begin{figure}[!h]
\begin{center}
\includegraphics[width=1\linewidth]{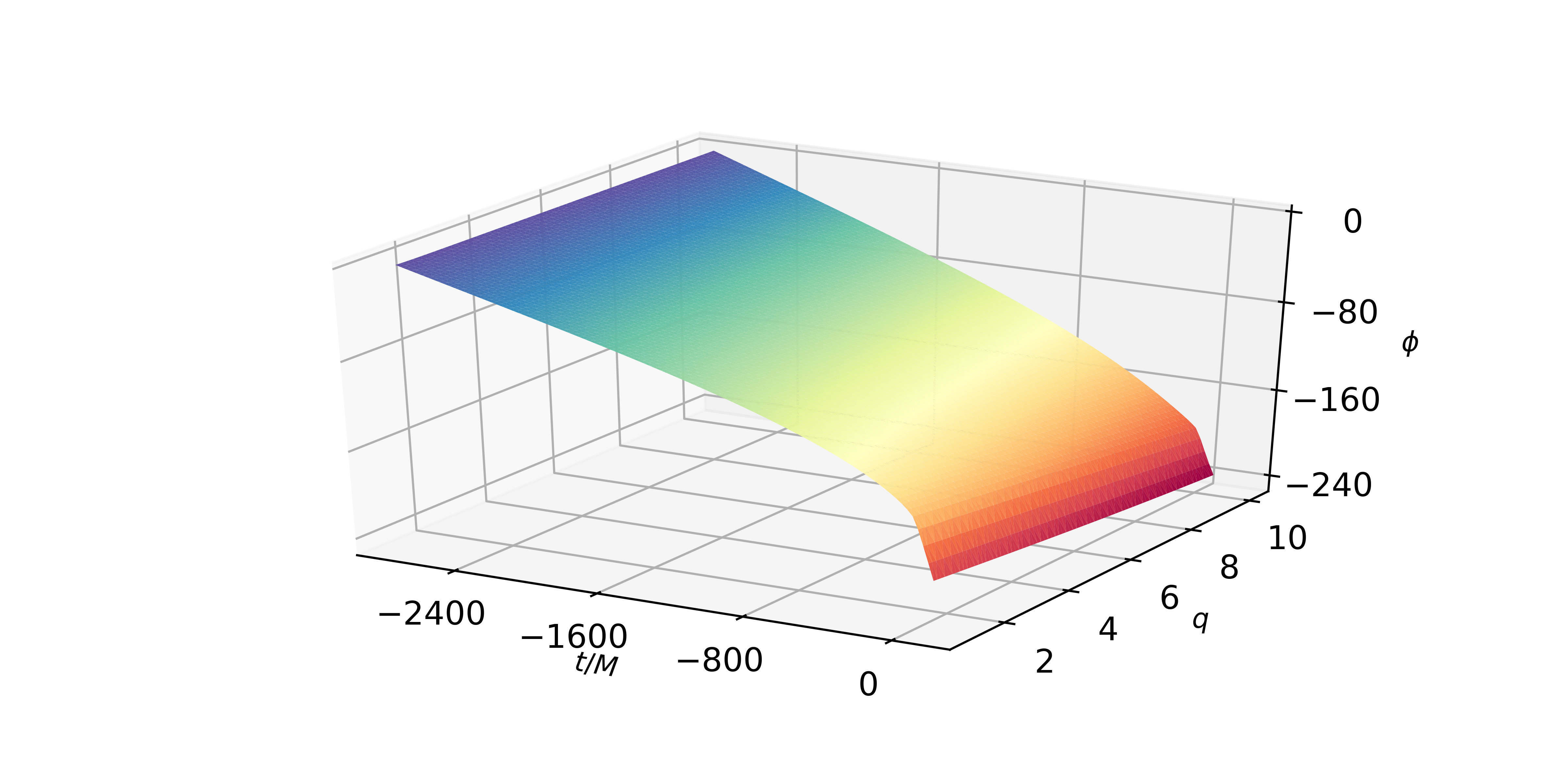}
\caption{Symbolic phase as a function of the mass ratio $q=m_1/m_2$ and time.}
\label{fig:phase}
\end{center}
\end{figure}

Although we were able to find high accuracy symbolic expressions for the phase in the considered interval of $q \in [1,10]$ (see Figure~\ref{fig:phase} for a symbolic model that is continuous in the whole interval), they resulted in large propagation errors in the reconstruction of the waveforms. The reason is different from phase {\em accumulation} errors in numerical relativity, since here we are dealing with global optimization errors and is simply the following: a change in phase $\phi \rightarrow \phi + \delta \phi$ in (\ref{complex}) leads to an error in the waveforms of the form
$$
h \rightarrow \tilde{h}:= A \times e^{i(\phi + \delta \phi)} \approx h\,(1+ i\delta \phi)  \, , 
$$ 
so $|\delta h|/A=\delta\phi$. In order to get a relative error of $1\%$ at least, we must have an order of  $0.01$ in the phase error $\delta \phi$. For the whole $q\in[1,10]$ phase symbolic model, in the results here obtained $\delta\phi$ is of order $1$ (with a relative error less than $10^{-2}$), leading to large errors when reconstructing the waveforms. 

A simple domain decomposition to solve this issue worked out for us: we subdivided the domain $q \in [1,10]$ into 9 equally spaced subdomains of the form
$$
q\in [1,2] \, , [2,3]  \, , \ldots \, ,  [8,9] \, ,  [9,10] \, ,
$$
We have not tried using a different number of domains, or of different relative sizes, since the point of this paper is to show how to build a sparse training set to avoid the high computational cost of symbolic regression through genetic programming, and that it works in practice in a highly non-trivial application. 

For simplicity, we have not imposed boundary conditions between the subdomain boundaries, although this could in principle be done. Domain decompositions are standard when solving partial differential equations (PDEs). In fact, it is possible that in more complex scenarios or even in this case a more elaborate scheme such as an hp-greedy domain decomposition \cite{Eftang:2011,Eftang:2010,Eftang2011,Eftang:2012:TCR:2197457.2197477} 
 (where the hp term is actually borrowed from domain decomposition and refinement in finite elements when solving partial PDEs) might be more efficient in terms of decreasing the number of subdomains and improve our results.  As an analogy, it is well known that when solving PDEs through a domain decomposition numerical solutions are {\em not} continuous across domain boundaries at fixed resolution (either in space, time, or both); this is usually addressed through {\it weak enforcement} of the solution across boundaries, usually in the form of penalty terms. For a detailed discussion on these topics in the context of Einstein's equations see, for example, \cite{Sarbach2012,Calabrese:2003vx,Lehner:2005bz}. In this article we focus on showing that, as a proof of concept, closed- and free-form expressions can be obtained without resorting to physical approximations. It might be possible to obtain accurate symbolic expressions without a domain decomposition, for example by enriching the dictionary through physically inspired functions, but this is left for future work. Here we focus on the basic elements of eliminating the curse of dimensionality in symbolic regression using modern approaches to reduced order modeling for parametrized systems.
  
When searching for symbolic expressions for the phase we used $20$ values in mass ratio and $285$ time nodes for each domain, with all points equally spaced. Due to the simple structure of the phase (see. Fig.~\ref{fig:phase}), subsampling was not necessary. 

Our highest accuracy results are the following: 

\beq
\begin{split}
&\phi_{[1,2]}(t,q)=b_1\,\text{atan2}(b_2,t)\,\text{sinh}^{-1}(b_3\,t) + \frac{b_4\,t\,\text{cosh}(q)\,\text{sinh}^{-1}(b_3\,t)}{\text{atan2}(b_2, t)}- b_5 - b_6\,t - b_7\,\text{cosh}(q) - b_8\,\text{sinh}^{-1}(t) - b_9\,\text{sinh}^{-1}(b_3\,t) 
\\
&b_1=16.9899198245249\,,\,b_2=0.0307964991839896\,,b_3=8.47135183989294\,,\,b_4=0.517828856434813\,,\\
&b_5=155.336099835965\,,\,b_6=23.6826537661638\,,b_7=1.67469018319234\,,\,b_8=26.3800439393647\,,\\
&b_9=64.7124400428524.
\end{split}
\eeq

\beq
\begin{split}
&\phi_{[2,3]}(t,q)=b_1\,t^2 + b_2\,t\,\text{e}^t - b_3 - b_4\,q- b_5\,t - b_6\,q\,t- b_7\,\text{sinh}^{-1}(b_8\,t) - b_9\,t\,\text{sinh}^{-1}(b_{8}\,t) \, , 
\\
&b_1=6.83585963818032\,,\,b_2=29.2184323756819\,,\,b_3=150.134280878147\,,\,b_4=5.63837416811418\,,\\
&b_5=336.187115348329\,,\,b_6=2.03916141015058\,,b_7=0.462051355197209\,,\,b_8=224.317310349048\,,\\
&b_9=42.1012107484649.
\end{split}
\end{equation*}

\beq
\begin{split}
&\phi_{[3,4]}(t,q)=b_1 - b_2\,q
- b_3\,t - b_4\,\text{sinh}^{-1}(t)- b_5\,q\,t
- b_6\,\text{tan}^{-1}(b_7\,t)
- b_8\,\text{tan}^{-1}(b_9\,t)\,\text{tan}^{-1}(b_{10}\,t) \, ,
\\
&b_1=-150.779871664876\,,\,b_2=5.48151705552273\,,
b_3=20.9244777358138\,,\,b_4=40.725476363926\,,\\
&b_5=1.98410145646586\,,\,b_6=58.0238952257731\,,\,
b_7=4.54076522147913\,,\,b_8=27.5236418372851\,,\\
&b_9=27.0569353214333\,,\,b_{10}=4.54076522147913.
\end{split}
\end{equation*}

\beq
\begin{split}
&\phi_{[4,5]}(t,q)=b_1\,\text{tanh}(t) + b_2\,t^2+ b_3\,q\,t\,\text{tanh}(t)- b_4-b_5\,q- b_6\,t - b_7\,\text{sinh}^{-1}(b_8\,t)- b_9\,t\,\text{sinh}^{-1}(b_{8}\,t) \, ,
\\
&b_1=8.5721194612964\,,\,b_2=4.77594286022371\,,\,
b_3=1.71392847899927\,,\,b_4=153.735830018958\,,\\
&b_5=4.65044447419562\,,\,b_6=301.17672677624\,,
b_7=0.401467421747743\,,\,b_8= 300.050625169173\,,\\
&b_9=34.6087658515619.
\end{split}
\eeq

\beq
\begin{split}
&\phi_{[5,6]}(t,q)=b_1\,\text{atan2}(b_2,\text{e}^{b_3\,t + (b_3\,t^2)^{b_4}})- b_5 - b_6\,q - b_7\,t- b_8\,\text{e}^t -b_9\,q\,t\,\text{atan2}(b_{10},\text{e}^{b_{3}\,t
+ (b_{3}\,t^2)^{b_{4}}}) \, ,
\\
&b_1=34.9967402431729\,,\,b_2=2.88396955316236\,,
b_3=18.5550015973867\,,\,b_4=0.56242721246326\,,\\
&b_5=144.334463413931\,,\,b_6=4.56653604042429\,,\,
b_7=33.8215169236648\,,\,b_8=53.5938957492291\,,\\
&b_9=1.05498674575652\,,\,b_{10}=0.0457688186271425.
\end{split}
\eeq

\beq
\begin{split}
&\phi_{[6,7]}(t,q)=b_1\,t^2 + b_2\,\text{sinh}^{-1}(b_3\,t)
+ b_4\,\text{atan2}(b_5, b_3\,t)+b_6\,t\,\text{sinh}^{-1}(b_3\,t)- b_7- b_8\,q - b_9\,t - b_{10}\,q\,t \, ,
\\
&b_1=1.16459728558136\,,\,b_2=2.04818853464955\,,
b_3=-3031.15606950987\,,\,b_4=5.63150899599863\,,\\
&b_5=5.54505004946533\,,\,b_6=19.5350382862564\,,\\
&b_7=164.847087315508\,,\,b_{8}=4.24649927087975\,,
b_9=239.435021215528\,,\,b_{10}=1.54209668558109.
\end{split}
\end{equation*}\\

\beq
\begin{split}
&\phi_{[7,8]}(t,q)=b_1\,\text{atan2}(b_2, t)
- b_3 - b_4\,q- b_5\,t- b_6\,t\,q - b_7\,\text{sinh}^{-1}(b_8\,t)
- b_9\,\text{tan}^{-1}(b_{10}\,t)\,\text{sinh}^{-1}(b_{11}\,t) \, ,
\\
&b_1=42.5234029423116\,,\,b_2=-0.454286929738614\,,
b_3=92.557645930759\,,\,b_4=3.87693080235069\,,\\
&b_5=21.6846366746129\,,\,b_6=1.40441380466267\,,\,
b_7=103.004616315544\,,\,b_8=3.4964623078196\,,\\
&b_9=31.6037083633658\,,\,b_{10}=19.0682642359228\,,\\
&b_{11}=3.4964623078196.
\end{split}\end{equation*}

\beq
\begin{split}
&\phi_{[8,9]}(t,q)=b_1\,t^2 + b_2\,\text{e}^{b_3\,t}+ b_4\,t\,\text{sinh}^{-1}(\text{sinh}^{-1}(b_5\,t)) - b_6 - b_7\,q - b_8\,t - b_9\,t\,q\,\text{sinh}^{-1}(\text{sinh}^{-1}(b_5\,t)) \, , 
\\
&b_1=0.223595449207837\,,\,b_2=0.0228895534195837\,,
b_3=53.3803388021468\,,\,b_4=108.64303226437\,,\\
&b_5=-24.5987724166689\,,\,b_6=162.16004867221\,,\,
b_7=3.50199154150564\,,\,b_8=307.395374115204\,,\\
&b_9=0.565057456771371.
\end{split}
\end{equation*}

\beq
\begin{split}
&\phi_{[9,10]}(t,q)=b_1\,\text{atan2}(t, b_2)+ b_3\,q\,t^2 - b_4- b_5\,q - b_6\,t- b_7\,t\,\text{sinh}^{-1}(b_8\,t)- b_9\,q^2\,\text{sinh}^{-1}(b_8\,t) \, ,
\\
&b_1=10.6554882562835\,,\,b_2=0.561000822409326\,,
b_3=0.390622102213681\,,\,b_4=164.868219149625\,,\\
&b_5=3.16971455362515\,,\,b_6=293.785912743941\,,\,
b_7=33.3325708174699\,,\,b_8=174.055527284276\,,\\
&b_9=0.00463931687452369.
\end{split}
\end{equation*}

\subsection*{Validation and accuracy of symbolic waveforms}

From the symbolic amplitude and phases we reconstructed the time series for the two polarizations of the GWs and compared them with the ground truth solutions using $10^5$ GWs per each subdomain $[q_i,q_{i+1}]$ $i=1, ..., 9$, and the whole $28,501$  time samples provided by {\it GWSurrogate}, leading to $\sim 10^6$ validation waveforms. 

\begin{figure}[!h]
\begin{center}
\includegraphics[width=0.7\linewidth]{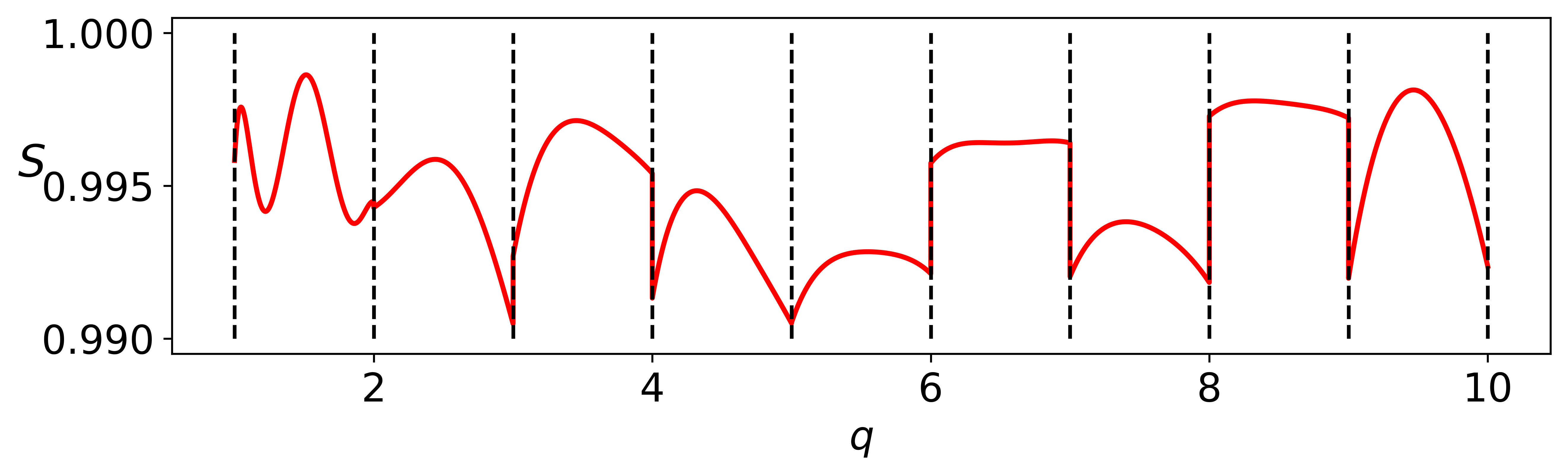}
\caption{Overlap $S(q)$ for the symbolic waveforms vs the mass ratio $q$, when compared to ground truth solutions, using $q \sim 10^6$ values. The minimum and maximum overlaps are $S=0.9905$ and $S=0.9986$, respectively. The dotted lines delimit each subdomain $[q_i,q_{i+1}]$ for $i=1 \ldots 9$.}
\label{fig:overlaps}
\end{center}
\end{figure}

The result is that the overlap $S=S(q)$ (\ref{eq:overlap}) in our approach gives values above $99\%$ for all cases. The main reason that we could do this a posteriori dense validation is due to the fact that ground truth solutions using surrogate models can be evaluated very quickly. The results are shown in Fig. \ref{fig:overlaps}. One should not reach any conclusion from the dependence of the overlap $S$ as a function of the mass ratio $q$, since these are {\em representations}, much as in domain decomposition approaches in NR (though usually in physical space, not in parameters).  For example, we could have chosen to show results for symbolic expressions with a more uniform error distribution, though it is worth emphasizing that the differences in the figure are in the order of $0.1\%$, which is below the accuracy required by Laser Interferometer GW detectors. Put differently, although not perhaps known for non-numerical relativists, this non-smooth behavior across boundaries is {\em always} present (though in a different sense) in numerical solutions to the Einstein equations when using state of the art domain decomposition or adaptive mesh refinement: continuity is only imposed within a given numerical resolution~\cite{Sarbach2012}. There are  {\em always} ``numerical jumps'' and at any fixed resolution the numerical solution to Eintein's equations is non-smooth; the idea is that these discontinuities are below an acceptable numerical error. The analogue in our approach is that across boundaries the symbolic phases are not smooth but they result in waveforms with overlaps with respect to the ground truth solutions which are below an acceptable error tolerance. As a remark, in GW science, this acceptable discrepancy is of $97 \%$, while our results guarantee at least $99 \%$.   

\begin{figure}[!h]
\begin{center}
\includegraphics[width=0.5\linewidth]{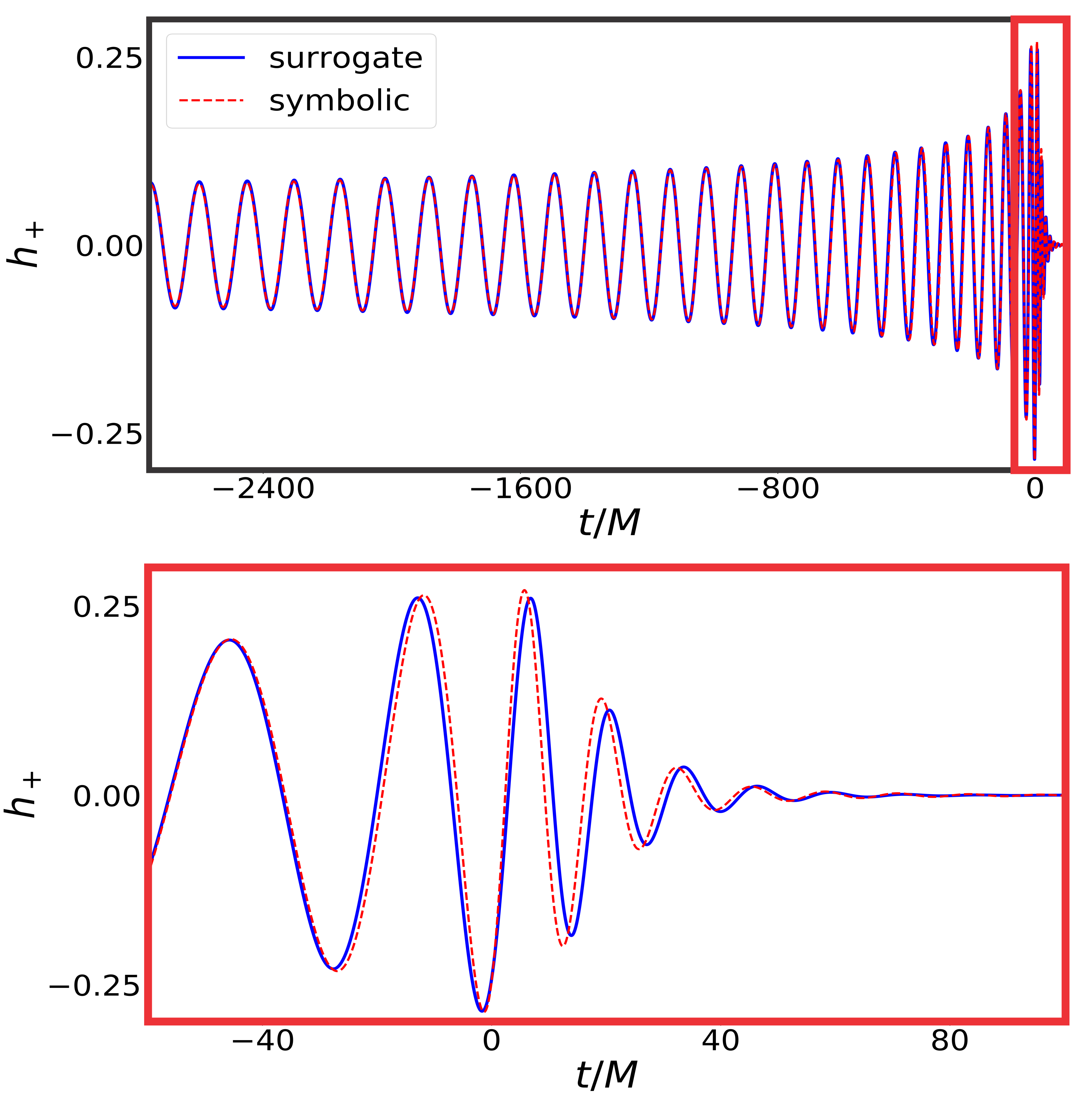}
\caption{Symbolic and surrogate waveforms corresponding to the {\em worst} match in a posteriori validation with respect to the surrogate model.  {\bf Top}: the whole waveforms, with the vertical redline prior to merger, at $t/M=-60$. {\bf Bottom}: zoom in of the waveforms in the range $t/M\in[-60,100]$. The differences near the merger are noticeable, but the overlap is still $S=0.9905$. }
\label{fig:worst}
\end{center}
\end{figure}

As an example, in Fig. \ref{fig:worst} we show the ground truth solution on top of its symbolic expression for $h_+$, corresponding to the {\em worst} match in the validation space for the whole interval $q \in [1,10]$. Results for $h_{\times}$ are similar since both modes are related simply by a $\pi/2$ phase difference.

\section*{Discussion} \label{sec:discussion}

In perspective, having high accuracy, closed-form (symbolic) expressions for the emitted gravitational waves as predicted by a theory as complex as Einstein's one of gravity, for a process as complex as the collision of two black holes, without any simplification in the theory (thus the ab initio-based emphasis), in a completely autonomous way, cannot be overemphasized.  

The standard procedure in GW phenomenological modeling is that one of ``stitching'' GWs from post-Newtonian (PN) expansions or Effective One Body approximations, ringdown, and a merger regime in between from numerical relativity in {\em some} way (there are many of them) and in this sense they do carry physical approximations and therefore are not ab initio-based in the sense here used. Similarly with hybrid models, where early in the inspiral stage some physical approximant is stitched to a numerical relativity solution. Here we have concentrated for training data on a model that is completely indistinguishable from high accuracy numerical relativity supercomputer simulations of the full Einstein equations. In this sense our results can be considered truly ab initio-based, since in the absence of exact solutions (except for those with high degrees of symmetry) high accuracy NR simulations of the Einstein equations are considered the true gold-standard.

Another point to emphasize is that, unlike most -- if not all -- phenomenological models, our closed-form expressions do not distinguish between inspiral, merger and ringdown, but model all regimes at once. One could have chosen to find symbolic expressions for the different regimes just mentioned, but in order to show the power of our approach we have chosen to model the whole inspiral-merger-ringdown case at once. The domain decomposition here presented is very simple. Having different models in parameter space is not unusual. In fact, it has allowed (among many other ingredients) to find new signals from public LIGO data~\cite{Venumadhav:2019lyq}.

Our approach is one of the many trends in the gravitational wave science community to incorporate tools from DS, ML and AI, but to our knowledge it is the first of its kind.  Because of this, and because genetic programming symbolic regression is meant to provide insights from data, it is difficult to anticipate the full impact and ramifications of our approach. It {\it might} be useful, for example, for other ones combining ROM with Deep Learning for GW inference~\cite{Chua:2018woh}, which produce, and start from closed-form expressions. 

We presented a proof of concept for a novel approach. A next natural, conceptually straightforward, step might be to apply it to the other multipole modes of~\cite{Blackman:2015pia}, and more complex systems such as the case of spinning, precessing black holes using, for example, the surrogate models of~\cite{PhysRevResearch.1.033015,PhysRevD.95.104023,PhysRevD.100.024002}. It is possible that for these cases, and higher dimensionality ones in parameter space in general, it would be beneficial when training GP-SR to use not only the EIM time nodes but also the greedy parameter values to increase sparseness and avoid the curse of dimensionality of SR searches. Another line of future research would be to use an hp-greedy refinement approach at the surrogate level to minimize the number of domains. The question of what is the optimal minimum number of subdomains and how it increases with parameter dimensionality is outside the scope of this work but remains as an interesting question. Making touch again with numerical relativity, the equivalent would be asking how the number of domains or levels of adaptive mesh refinement changes with resolution. Even in an established field such as NR, where there is little to no theory for equations as complicated as the Einstein ones, one usually proceeds through numerical experiments.

Even though here we have focused on symbolic expressions based on surrogates built from high accuracy numerical relativity simulations, our approach can be applied to other surrogates based on RB and the EIM, for example those based on EOB or PN ones~\cite{Field:2013cfa,Lackey:2016krb}. 

The sparse yet near-optimal subsampling in time using the EIM is a key ingredient in our approach for finding closed-form expressions, so it is not clear that other surrogate models (say based on Gaussian regression, see, e.g.Ref.~\cite{Lackey:2018zvw}) can do so. There might be potential in enriching the dictionary here used for GP-SR (composed of elementary functions and basic arithmetic operations) using phenomenological or other physically based symbolic models. In this sense, using GP-SR should outperform any other kind of physics-based fits by design being free-form. Also, any other fits can be used as a bootstrap to enrich the dictionary of GP-SR; this can also be done as more dimensions and physics complexity are added~\cite{Schmidt03042009}. 

Anyone willing to qualitatively reproduce or extend the results of this paper can apply for a free academic license of DataRobot~\cite{datarobot}, an Automated Machine Learning framework, which integrates the GP algorithm used in this work. Our results can be easily accesible through a Jupyter notebook at https://github.com/aaronuv/SymbolicGWs.

In general terms, our approach should be applicable to other disciplines beyond gravitational wave science since computational complexity is a common problem in genetic programming. 

\bibliography{references}

\section*{Acknowledgements}

The main ideas of this project were initiated with Rory Smith, to whom we are extremely thankful, as well as Alan Weinstein and Yanbei Chen for hospitality at Caltech. We thank Hod Lipson for introducing us to symbolic regression, the team at Nutonian for academic licenses of Eureqa (now part of DataRobot) and a steep discount on their enterprise parallel version, and Jorge Pullin for comments on a previous draft of this paper. We also thank two anonymous referees for valuable comments and feedback on a previous version of this manuscript. 
This project was supported in part by CONICET (Argentina).

\section*{Author contributions statement}
Manuel Tiglio and Aar\'on Villanueva contributed equally to the research leading to this paper, as well as its writing and review. The list of authors has been ordered alphabetically. 

\end{document}